\documentclass[aps,prl,reprint,amsmath,amssymb]{revtex4-1}
\usepackage{graphicx}
\usepackage{subfigure}
\usepackage{epstopdf}
\usepackage{color}
\usepackage{multirow}
\usepackage{setspace}
\usepackage{overpic}
\usepackage{makecell}
\usepackage[bookmarksnumbered, pdfstartview=FitH,colorlinks,urlcolor=blue, citecolor=blue,linkcolor=blue] {hyperref}
\usepackage{lineno}
\usepackage{bm}
\usepackage{float}
\usepackage{rotating}
\usepackage[utf8]{inputenc}
\usepackage[separate-uncertainty=true,table-align-uncertainty=true,]{siunitx}

\let\oldequation\equation
\let\oldendequation\endequation
\renewenvironment{equation}
 {\linenomathNonumbers\oldequation}
 {\oldendequation\endlinenomath}

\newcommand{\klnu}{D^{0(+)}\to \bar K\ell^+\nu_{\ell}}

\newcommand{\kenu}{D^0\to K^-e^+\nu_e}
\newcommand{\ffK}{f_+(0)}
\newcommand{\kmunu}{D^0\to K^-\mu^+\nu_\mu}
\newcommand{\ksenu}{D^+\to \bar K^0e^+\nu_e}
\newcommand{\ksmunu}{D^+\to \bar K^0\mu^+\nu_\mu}

\newcommand{\BESIIIorcid}[1]{\href{https://orcid.org/#1}{\hspace*{0.1em}\raisebox{-0.45ex}{\includegraphics[width=1em]{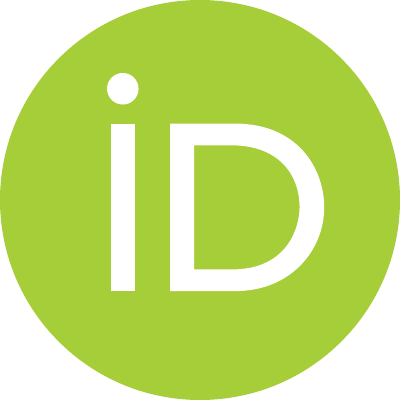}}}}

\begin{document}
\title{\boldmath First Experimental Constraint on the Scalar Current in the $\klnu$ Transition}
\author{
M.~Ablikim$^{1}$\BESIIIorcid{0000-0002-3935-619X},
M.~N.~Achasov$^{4,c}$\BESIIIorcid{0000-0002-9400-8622},
P.~Adlarson$^{81}$\BESIIIorcid{0000-0001-6280-3851},
X.~C.~Ai$^{87}$\BESIIIorcid{0000-0003-3856-2415},
C.~S.~Akondi$^{31A,31B}$\BESIIIorcid{0000-0001-6303-5217},
R.~Aliberti$^{39}$\BESIIIorcid{0000-0003-3500-4012},
A.~Amoroso$^{80A,80C}$\BESIIIorcid{0000-0002-3095-8610},
Q.~An$^{77,64,\dagger}$,
Y.~H.~An$^{87}$\BESIIIorcid{0009-0008-3419-0849},
Y.~Bai$^{62}$\BESIIIorcid{0000-0001-6593-5665}
% ...
\\
\vspace{0.2cm}
(BESIII Collaboration)\\
\vspace{0.2cm} {\it
$^{1}$ Institute of High Energy Physics, Beijing 100049, People's Republic of China\\
% ...
$^{4}$ Budker Institute of Nuclear Physics SB RAS (BINP), Novosibirsk 630090, Russia\\
% ...
$^{31}$ INFN Sezione di Ferrara, (A)INFN Sezione di Ferrara, I-44122, Ferrara, Italy; (B)University of Ferrara, I-44122, Ferrara, Italy\\
% ...
$^{39}$ Johannes Gutenberg University of Mainz, Johann-Joachim-Becher-Weg 45, D-55099 Mainz, Germany\\
% ...
$^{62}$ Southeast University, Nanjing 211100, People's Republic of China\\
% ...
$^{64}$ State Key Laboratory of Particle Detection and Electronics, Beijing 100049, Hefei 230026, People's Republic of China\\
% ...
$^{77}$ University of Science and Technology of China, Hefei 230026, People's Republic of China\\
% ...
$^{80}$ University of Turin and INFN, (A)University of Turin, I-10125, Turin, Italy; (B)University of Eastern Piedmont, I-15121, Alessandria, Italy; (C)INFN, I-10125, Turin, Italy\\
$^{81}$ Uppsala University, Box 516, SE-75120 Uppsala, Sweden\\
% ...
$^{87}$ Zhengzhou University, Zhengzhou 450001, People's Republic of China\\
% ...
$^{\dagger}$ Deceased\\
% ...
$^{c}$ Also at the Novosibirsk State University, Novosibirsk, 630090, Russia\\
% ...
}
}

\begin{abstract}

Using 20.3 fb$^{-1}$ of $e^+e^-$ collision data taken at the
center-of-mass energy of $\sqrt{s}=3.773$ GeV, we report the first
experimental constraint on the scalar current in the $\klnu$
transitions, based on a simultaneous fit to the first measured
forward-backward asymmetries and precisely determined partial decay
rates. The parameters of the scalar current are determined to be ${\rm
Re}(c_S^{\mu})=0.017\pm0.008_{\rm stat}\pm0.006_{\rm syst}$ and
${\rm Im}(c_S^{\mu})=\pm(0.077\pm0.011_{\rm stat}\pm0.009_{\rm
syst})$, which deviates from the SM by a significance of
$2.3\sigma$. The branching fractions of $D^{0(+)}\to \bar
K\ell^+\nu_\ell$, the hadronic form factor $f_+(0)$, and the modulus
of the $c\to s$ CKM matrix element $|V_{cs}|$ are also determined with
improved precision. In addition, lepton flavor universality is tested
via the ratios of the decay rates between semi-muonic and
semi-electronic decays in the full momentum
transfer range and in subranges.

\end{abstract}

\maketitle

In the Standard Model~(SM), the couplings between the gauge bosons and
the three families of leptons are expected to be the same, known as
lepton flavor universality~(LFU). Any observation of LFU violation
indicates potential new physics (NP) effects beyond the SM. In recent
years, several hints of lepton non-universal effects have been
reported in the $B$ meson semileptonic decays via $b\to
c\tau^+\nu_{\tau}$~\cite{BaBar:2012obs,Belle:2015qfa,LHCb:2023zxo,LHCb:2023uiv,LHCb:2024jll} and
$b\to s
\ell^+\ell^-$~($\ell=e,\mu$)~\cite{LHCb:2013ghj,Belle:2016fev,CMS:2017rzx,ATLAS:2018gqc,LHCb:2024onj}
transitions. While some of these violations disappear after further update measurements, some anomalies remain. Various theoretical scenarios involving two-Higgs-doublet
models~\cite{Iguro:2022uzz,Blanke:2022pjy} and leptoquark
models~\cite{Sakaki:2013bfa,Becirevic:2018afm}, as well as model
independent investigations~\cite{Iguro:2024hyk}, suggest that the
violations could be attributed to a scalar current in the weak interaction. Given the weak interaction in the SM is only contributed by vector and axial-vector currents, such a scalar current would be a clear sign of NP. While constrained in $B$ meson decays~\cite{Iguro:2024hyk,LHCb:2024onj}, its magnitude in charm decays remains virtually unexplored. References~\cite{Barranco:2013tba,Fajfer:2015ixa,Barranco:2016njc,Zhang:2018jtm,Bolognani:2024cmr}
state that the scalar current could also cause a deviation from the SM
in the $c\to s\ell^+\nu_\ell$ transition. The partial decay rates
$\frac{d\Gamma}{dq^2}$ and the forward-backward asymmetries $A_{\rm
FB}(q^2)$ of the semileptonic~(SL) decays $\klnu$ are sensitive to the
scalar current contributions in the $c\to s\ell^+\nu_\ell$
transition. Therefore, experimental studies of these decays are
important to explore the parameter space of the scalar current, which
may reveal unknown scalar boson contributions beyond the
SM~\cite{Iguro:2022uzz,Blanke:2022pjy,Sakaki:2013bfa,Becirevic:2018afm}.

In addition, precise measurements of the $\klnu$ decay dynamics are
critical to determine the hadronic transition form factor~(FF) $\ffK$
and the modulus of the $c\to s$ Cabibbo-Kobayashi-Maskawa~(CKM) matrix
element $|V_{cs}|$. A precisely measured $\ffK$ allows for stringent
tests of different theoretical
predictions~\cite{Lubicz:2017syv,Khlopov:1978id,Gershtein:1976mv,Chakraborty:2021qav,Parrott:2022rgu,FermilabLattice:2022gku,Wu:2006rd,Verma:2011yw,Ivanov:2019nqd,Faustov:2019mqr,Ke:2023qzc},
and an accurately determined $|V_{cs}|$ is important for testing the
CKM matrix unitarity. Also, the full and partial decay widths
$\Delta\Gamma_{\ell=e,\mu}$ of $\klnu$, measured with improved precision, provide an important test of the LFU via the ratio
$\mathcal{R}_{\mu/e}\equiv\frac{\Delta\Gamma_\mu}{\Delta\Gamma_e}$.

Previously, the branching fraction~(BF) and the decay dynamics of
$\klnu$ have been widely investigated by
BES~\cite{BES:2004rav,BES:2004obp,BES:2006kzp},
BaBar~\cite{BaBar:2007zgf}, Belle~\cite{Belle:2006idb},
CLEO~\cite{CLEO:2005rxg,CLEO:2005cuk,CLEO:2007ntr,CLEO:2009svp}, and
BESIII~\cite{BESIII:2021mfl,BESIII:2015tql,BESIII:2018ccy,BESIII:2017ylw,BESIII:2016hko,BESIII:2015jmz,BESIII:2016gbw,BESIII:2024slx}. However,
no experimental study of the scalar current and the forward-backward
asymmetry of $\klnu$ has been reported yet. With 20.3 fb$^{-1}$ of
$e^+e^-$ collision data taken at the center-of-mass energy of
$\sqrt{s}=3.773$ GeV with the BESIII detector, this Letter presents
the first experimental constraint on the scalar current in the $\klnu$
transition, based on a simultaneous fit to the first measured
forward-backward asymmetries and precisely determined partial decay
rates. The BFs of $\klnu$, as well as the product of the FF $f_+(0)$
and the modulus of the CKM matrix element $|V_{cs}|$ are determined
with improved precision. The results supersede those reported
in Ref.~\cite{BESIII:2024slx}, which was based on a portion of the data
sample used in this Letter. Charge conjugate modes are implied
throughout this Letter.


Details about the design and performance of BESIII detector are given
in Refs.~\cite{BESIII:2009fln,Huang:2022wuo}. The inclusive Monte
Carlo~(MC) samples, described in
Refs.~\cite{GEANT4:2002zbu,Jadach:1999vf,Jadach:2000ir,Lange:2001uf,Ping:2008zz,Golonka:2005pn},
are used to estimate the background in this analysis. The signal MC
samples of $\klnu$ are simulated with a specific two-parameter series
expansion model~\cite{Becher:2005bg} using parameters obtained in this
Letter.


At $\sqrt{s}=3.773$ GeV, the $D$ and $\bar{D}$ mesons are produced in
pairs without accompanying particles in the final state. This property
allows us to study SL $D$ decays with the double-tag~(DT)
method~\cite{MARK-III:1985hbd}. First, the single-tag~(ST) $\bar{D}$
mesons are reconstructed with the hadronic decay modes $\bar{D}^0\to
K^+\pi^-$, $K^+\pi^-\pi^0$, $K^+\pi^-\pi^-\pi^+$, $K_S^0 \pi^+\pi^-$,
$K^+\pi^-\pi^0\pi^0$, and $K^+\pi^-\pi^-\pi^+\pi^0$; $D^-\to
K^+\pi^-\pi^-$, $K_S^0\pi^-$, $K^+\pi^-\pi^-\pi^0$, $K_S^0\pi^-\pi^0$,
$K_S^0\pi^+\pi^-\pi^-$, and $K^+K^-\pi^-$. In the presence of the ST
$\bar{D}$ mesons, candidates for the signal decays $\klnu$ are
selected to form DT events, in which $\bar{K}^0$ is reconstructed via
$K^0_S\to\pi^+\pi^-$ and the lepton $\ell$ denotes the electron and
muon. The BF of the signal decay is determined by \begin{equation}
\label{eq:BF} \mathcal{B}_{\rm sig} = \frac{N_{\rm DT}}{N^{\rm
tot}_{\rm ST}\cdot\bar{\epsilon}_{\rm sig}\cdot\mathcal{B}_{\rm sub}}.
\end{equation} Here, $N^{\rm tot}_{\rm ST}=\sum_{i}{N^i_{\rm ST}}$ and
$N_{\rm DT}$ are the total ST and DT yields after summing over all tag
modes; $\bar{\epsilon}_{\rm sig}=\sum_{i}\frac{N^i_{\rm ST}}{N^{\rm
tot}_{\rm ST}}\frac{\epsilon^i_{\rm DT}}{\epsilon^i_{\rm ST}}$ is the
averaged signal efficiency of selecting $\klnu$ in the presence of ST
$\bar{D}$ mesons, where $\epsilon^i_{\rm ST}$ and $\epsilon^i_{\rm
DT}$ are the ST and DT efficiencies for the $i$-th tag mode,
respectively; $\mathcal{B}_{\rm sub}=\mathcal{B}(\bar{K}^0\to K^0_S,
K^0_S\to\pi^+\pi^-)$~\cite{ParticleDataGroup:2024cfk} for $D^+$
channels and $\mathcal{B}_{\rm sub}=1$ for $D^0$ channels.

The selection criteria of $\pi^\pm,K^\pm,K_S^0,\gamma$, and $\pi^0$
candidates are the same as those in
Ref.~\cite{supplemental_material}. Two kinematic variables, the energy
difference $\Delta E \equiv E_{\bar{D}}-E_{\rm beam}$ and the
beam-constrained mass $M_{\rm BC}\equiv \sqrt{E_{\rm
beam}^2-|\vec{p}_{ \bar{D}}|^2}$, are used to distinguish the ST
$\bar{D}$ mesons from combinatorial backgrounds, where $E_{\rm beam}$
is the beam energy and $(E_{\bar{D}},\vec{p}_{ \bar{D}})$ is the
four-momentum of the ST $\bar{D}$ in the $e^+e^-$ rest frame. The
combination with the smallest $|\Delta E|$ is chosen if there are
multiple combinations in an event.

The ST candidates must satisfy tag mode requirements for
$\Delta E$, corresponding to about $\pm 3.5\sigma$ around the fitted
peaks.  The ST yields in data for each tag mode are obtained by
fitting individual $M_{\rm BC}$ distributions. Candidates with $M_{\rm
BC}\in [1.859,1.873]$~GeV/$c^2$ for $\bar D^0$ and $M_{\rm BC}\in
[1.863,1.877]$~GeV/$c^2$ for $D^-$ are retained for further
analysis~\cite{BESIII:2018nzb}. The $\Delta E$ requirements, the ST
yields in data, and the ST efficiencies for different tag modes are
provided in Ref.~\cite{supplemental_material}. Summing over all the tag
modes, $N^{\rm tot}_{\rm ST}$ is determined to be
$(10~646.9\pm3.8)\times10^3$ for $D^-$ and
$(20~226.3\pm5.5)\times10^3$ for $\bar{D}^0$.

In the presence of the tagged $\bar{D}$, candidates for $\kenu$,
$\kmunu$, $\ksenu$, and $\ksmunu$ are selected from the residual
tracks and showers, with the same selection criteria as those in
Ref.~\cite{BESIII:2024slx} except that the $M_{\bar{K}\mu}$
requirements are optimized to $M_{K^-\mu^+}<1.56$ GeV/$c^2$ for
$\kmunu$ and $M_{K_S^0\mu^+}<1.59$ GeV/$c^2$ for $\ksmunu$. The
detailed descriptions are available in
Ref.~\cite{supplemental_material}.

Since the neutrino is undetectable by the BESIII detector, its
four-momentum $(E_{\rm miss},\vec{p}_{\rm miss})$ is obtained by
calculating the missing energy and momentum, defined as $E_{\rm
miss}\equiv E_{\rm beam}-E_{\bar{K}}-E_{\ell^+}$ and $\vec{p}_{\rm
miss}\equiv \vec{p}_{D}-\vec{p}_{\bar{K}}-\vec{p}_{\ell^+}$, where
$(E,\vec{p})$ is the four-momentum of $\bar{K}$ or $\ell^+$. A
kinematic variable $U_{\rm miss}\equiv E_{\rm miss} - |\vec{p}_{\rm
miss}|$ is defined to obtain the signal yield. To further improve the
resolution of the neutrino, the momentum of the $D$ meson is constrained
as $\vec{p}_{D}=-\hat{p}_{\bar{D}}\sqrt{E^2_{\rm
beam}-m^2_{\bar{D}}}$, where $\hat{p}_{\bar{D}}$ is the unit vector in
the momentum direction of the ST $\bar{D}$ meson and $m_{\bar{D}}$ is
the known $\bar{D}$ mass~\cite{ParticleDataGroup:2024cfk}.

\begin{figure}[htbp]
	\centering
	\includegraphics[width=8.7cm]{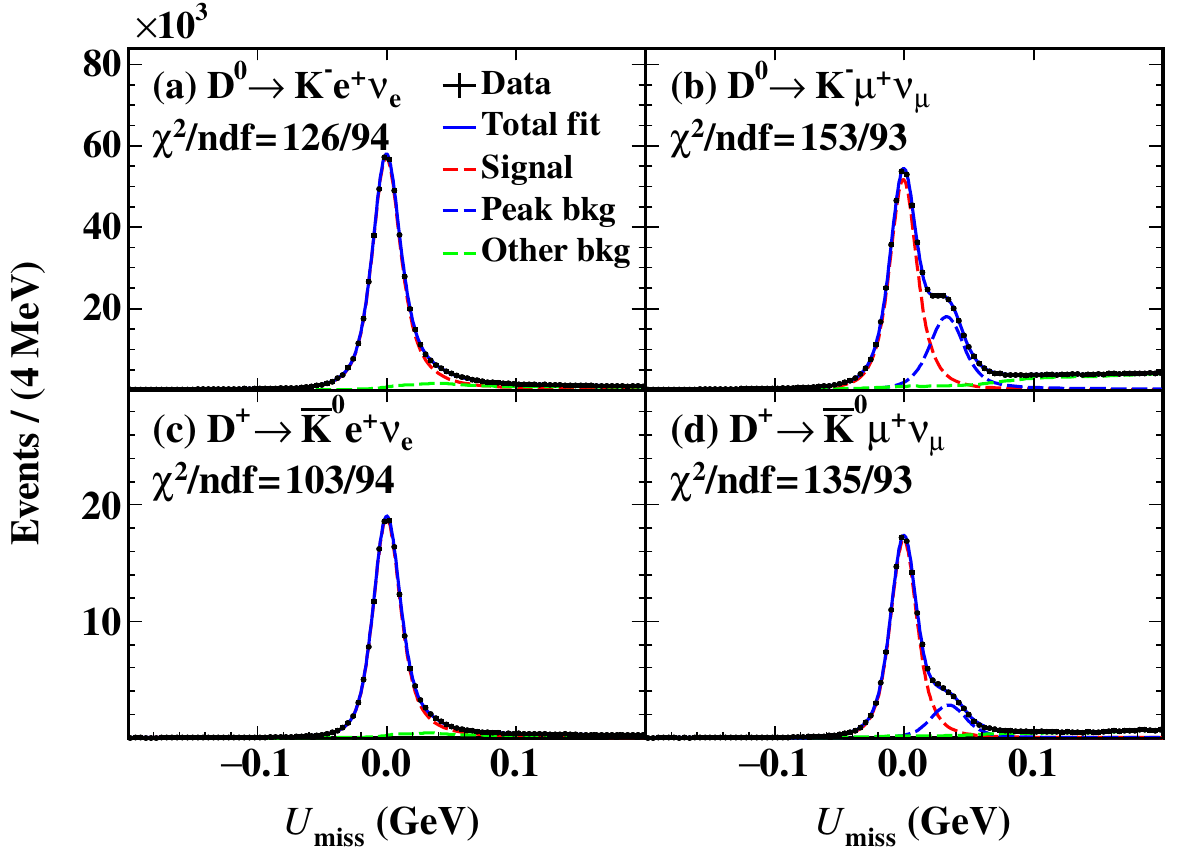}
	\caption{Fits to the $U_{\rm miss}$ distributions of $\klnu$. The points with error bars are data. The blue solid lines denote the total fits. The red, blue, and black dashed lines show the signal, peaking background, and other background contributions, respectively.}
	\label{fig:fit_umiss}
\end{figure}

The signal yields of $\klnu$ are determined from binned maximum
likelihood fits to the $U_{\rm miss}$ distributions of the accepted
candidates, as shown in Fig.~\ref{fig:fit_umiss}. In the fits, the
signals are modeled by the MC-simulated shape convolved with Gaussian resolution functions with free parameters to account for the
difference between data and MC simulation. The peaking backgrounds
$D\to \bar{K}\pi^+\pi^0$ for the $D\to \bar{K}\mu^+\nu_\mu$ are
described with the simulated shape convolved with the same Gaussian
functions with floating yields. The other background components are
modeled by the inclusive MC sample. Both the yields of signal and
backgrounds are floated in the fits. The fitted signal yields in data,
the averaged signal efficiencies corrected for the data-MC
differences~\cite{supplemental_material} on tracking and PID
efficiencies, and the BFs are summarized in
Table~\ref{tab:Bfs}. Input and output checks have been performed with MC simulation to ensure the analysis workflow are correct. Based on these, we determine the ratios
of the BFs between semi-muonic and semi-electronic $D$ decays:
$\mathcal{R}^{D^0}_{\mu/e}=0.971\pm0.003_{\rm stat}\pm0.004_{\rm
	syst}$ and $\mathcal{R}^{D^+}_{\mu/e}=0.982\pm0.004_{\rm
	stat}\pm0.002_{\rm syst}$, which are consistent with the LFU
predicted value of $0.975\pm0.001$~\cite{Riggio:2017zwh} within
1.5$\sigma$.

\begin{table}[htbp]
	\setlength{\abovecaptionskip}{0.0cm}
	\centering
	\caption{The fitted signal yields in data ($N_{\rm DT}$), the
	averaged signal efficiencies corrected for data-MC
	differences ($\bar{\epsilon}_{\rm sig}$), and the
	BFs ($\mathcal{B}_{\rm sig}$), where the first uncertainties
	are statistical and the second systematic.}
	\begin{footnotesize}
	\scalebox{1.0}{
	\begin{tabular}{cccc}
		\hline\hline
		Channel& $N_{\rm DT}$~($\times10^3$) &$\bar{\epsilon}_{\rm sig}$~(\%)&$\mathcal{B}_{\rm sig}~(\%)$ \\
		\hline
		$\kenu$ & $489.8\pm0.8$ &$68.25\pm0.02$ &$3.548\pm0.006\pm0.017$ \\
		$\kmunu$ & $403.9\pm0.8$ &$57.96\pm0.02$&$3.445\pm0.007\pm0.017$\\
		$\ksenu$ & $149.3\pm0.4$ & $45.39\pm0.02$ & $8.928\pm0.025\pm0.050$ \\
		$\ksmunu$ & $125.1\pm0.4$ & $38.74\pm0.02$ & $8.770\pm0.029\pm0.053$ \\
		\hline\hline
	\end{tabular}
	}
	\end{footnotesize}
	\label{tab:Bfs}
\end{table}

The systematic uncertainties in the BF measurements include the
uncertainties due to ST yields~(0.3\%), $K_S^0$
reconstruction~(0.4\%), correction factors for data-MC differences on
tracking~(0.1\%) and PID~(0.1\%) efficiencies,
requirements on maximum energy of the extra photon~(0.1\%) and
$M_{K\ell}$~(negligible), $U_{\rm miss}$ fit~(0.08$\sim$0.29\%),
quoted $\mathcal{B}_{\rm sub}$~(0.07\%), MC statistics~(0.04\%), and
MC model~(0.04$\sim$0.15\%). The detailed descriptions and values are
available in Ref.~\cite{supplemental_material}.

\begin{figure*}[htbp]
	\centering
	\includegraphics[width=18.0cm]{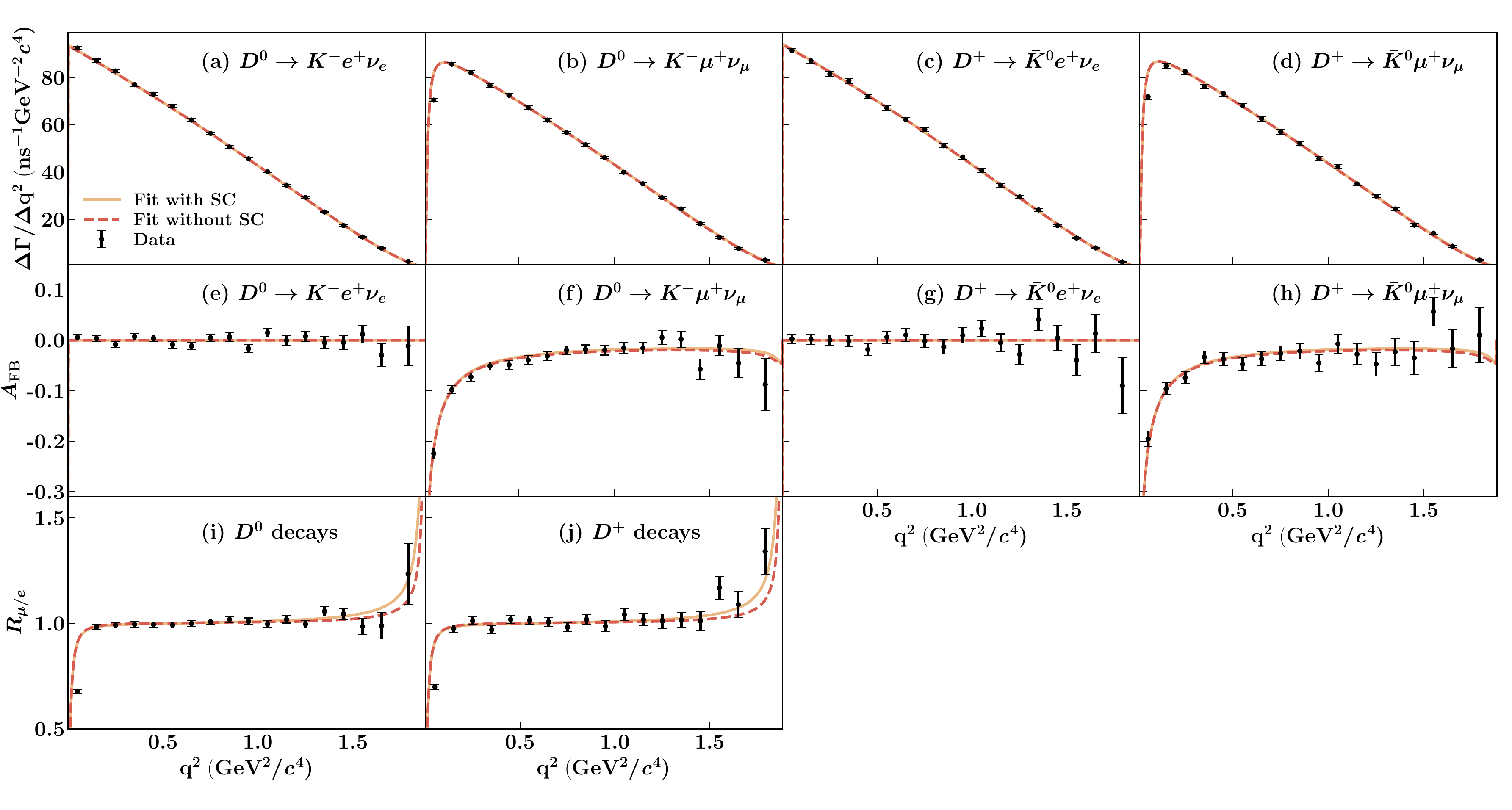}

	\caption{Simultaneous fit to the measured partial decay
	rates~(top) and forward-backward asymmetries~(middle) of
	$\klnu$, and the fit projections on the ratios of differential
	decay rates $\mathcal{R}_{\mu/e}$~(bottom). The black points
	with red error bars are data, the solid lines are the fit
	projections with scalar current~(SC) contribution included,
	and the dashed lines are the fit projections without scalar
	current contribution.}

	\label{fig:combined_fit}
\end{figure*}

The $q^2$-binned partial decay rates in the $i$-th $q^2$ interval
$\Delta \Gamma_{i} = \int^{q^2_{{\rm max}(i)}}_{q^2_{{\rm
min}(i)}}{\frac{d\Gamma}{dq^2} dq^2}$ are determined using $\Delta
\Gamma_{i}=N^{i}_{\rm prd}/(\tau_{D}\cdot N^{\rm tot}_{\rm ST})$,
where $N^{i}_{\rm prd}$ is the number of produced events in the $i$-th
$q^2$ interval and $\tau_{D}$ is the $D$
lifetime~\cite{ParticleDataGroup:2024cfk}. $N^{i}_{\rm
prd}=\sum_{j}(\epsilon^{-1}_{ij}) N^{j}_{\rm obs}$ is obtained by
performing similar fits to the $U_{\rm miss}$ distributions in each
$q^2$ interval to determine the numbers of observed events $N^{j}_{\rm
obs}$ and then transforming them into the produced ones with the
efficiency matrix $\epsilon_{ij}$, based on MC
simulation~\cite{supplemental_material}. The systematic uncertainty
studies are performed with similar approaches as those for the BF
measurement with details shown in
Ref.~\cite{supplemental_material}. The measured partial decay rates of
$\klnu$ as well as the calculated ratios $\mathcal{R}_{\mu/e}$ are
shown in the top and bottom figures of Fig.~\ref{fig:combined_fit},
respectively.

The $q^2$-binned forward-backward asymmetry in the $i$-th $q^2$
interval
\begin{equation} A_{{\rm FB},i}=\int^{q^2_{{\rm
max}(i)}}_{q^2_{{\rm min}(i)}}{A_{{\rm FB}}(q^2)\frac{d\Gamma}{dq^2}
dq^2}\bigg{/}\int^{q^2_{{\rm max}(i)}}_{q^2_{{\rm
min}(i)}}{\frac{d\Gamma}{dq^2} dq^2} \end{equation}
is measured by 
$A_{{\rm FB},i} = \frac{N^{i}_{\rm prd,f}-N^{i}_{\rm prd,b}}{N^{i}_{\rm
prd,f}+N^{i}_{\rm prd,b}}$, where $N^{i}_{\rm prd,(f,b)}$ is the
number of produced forward~($\cos\theta_W>0$) and
backward~($\cos\theta_W<0$) events in the $i$-th $q^2$ interval
obtained using the similar approach as that of partial decay rates, where
$\theta_W$ is the angle between the flight direction of lepton and
opposite flight direction of $D$ in the $\ell\nu_\ell$ rest
frame. Some systematic uncertainties cancel, except for the $U_{\rm
miss}$ fit~($(1\sim8)\times10^{-4}$), correction factors related to
lepton~($1\times10^{-4}$), requirements on $M_{K\ell}$~(negligible),
and MC statistics~($<7\times10^{-4}$), where the values correspond to
absolute uncertainties. The measured forward-backward asymmetries of
$\klnu$ are shown in Fig.~\ref{fig:combined_fit}.

The scalar current contribution is obtained via a simultaneous fit to
the partial decay rates $\frac{d\Gamma}{dq^2}$ and forward-backward
asymmetries $A_{\rm FB}(q^2)$, the explicit expressions of which are
written as~\cite{Fajfer:2015ixa,Faustov:2019mqr}
\begin{equation}
\small
\frac{d\Gamma}{dq^2} = \mathcal{N}(q^2)\left(1-\frac{m_\ell^2}{q^2}\right)^2\left[\left(1+\frac{m_\ell^2}{2q^2}\right)|h_0(q^2)|^2 + \frac{3m_\ell^2}{2q^2}|h_t(q^2)|^2\right]
\end{equation}
and
\begin{equation}
\small
A_{\rm FB}(q^2)=\frac{3\mathcal{N}(q^2)}{2}\frac{1}{d\Gamma^{\ell}/dq^2}\left(1-\frac{m_\ell^2}{q^2}\right)^2\frac{m_\ell^2}{q^2} {\rm Re}\left(h_0(q^2)h^*_t(q^2)\right),
\end{equation}
respectively. Here,
$\mathcal{N}(q^2)=\frac{G_F^2|V_{cs}|^2|\textbf{q}|q^2}{96\pi^3
m^2_D}$ is the factor including the Fermi coupling constant $G_F$, the
CKM matrix element $V_{cs}$, and the mass of $D$ meson $m_{D}$; $q$ is
the four momentum of the $\ell\nu_\ell$ system in the $D$ rest frame
and $|\textbf{q}|$ is the magnitude of the three momentum of $q$;
$m_\ell$ is the mass of the lepton. With scalar currents included, the
hadronic helicity amplitudes $h_{0(t)}(q^2)$ are modified as
\begin{equation}
\begin{split}
\label{eq:helicity_amp}
h_0(q^2) &= \frac{2 m_D |\textbf{q}|}{\sqrt{q^2}}f_{+}(q^2),\\
h_t(q^2) &= \left(1+c_S^\ell\frac{q^2}{m_\ell(m_s-m_c)}\right)\frac{m_D^2-m_K^2}{\sqrt{q^2}}f_0(q^2),
\end{split}
\end{equation}
where $f_{+,0}(q^2)$ are the form factors~(FFs); $c_S^\ell=c^{\ell}_R+c^{\ell}_L$ is the scalar combination of Wilson coefficients for the right and left-handed scalar currents~\cite{Fajfer:2015ixa}; $m_c=1.273$ GeV/$c^2$ and $m_s=93.5$ MeV/$c^2$~\cite{ParticleDataGroup:2024cfk} are the $\overline{\rm MS}$ masses at the energy scale $\mu=m_c$ and 2 GeV, respectively; $m_K$ is the mass of kaon. The FF $f_{+}(q^2)$ is parametrized in the two-parameter series expansion approach~\cite{Becher:2005bg} as
\begin{equation}
\label{equation:ffp}
\begin{array}{l}
\displaystyle f_{+}\left(q^2\right)=\frac{1}{P\left(q^2\right)\Phi\left(q^2\right)}\frac{f_{+}\left(0\right)P\left(0\right)\Phi\left(0\right)}{1+r_{1}
	\left(t_{0}\right)z\left(0,t_{0}\right)}\\
\displaystyle\times\left(1+r_{1}\left(t_{0}\right)\left[z\left(q^2,t_{0}\right)\right]\right),
\end{array}
\end{equation}
where the parameters $r_{1}\left(t_{0}\right)$ and $f_{+}(0)$ are
floated in the fit. The FF $f_{0}(q^2)$ is modeled with the one-parameter
series expansion as
\begin{equation}
f_0(q^2) = \frac{1}{P(q^2)\Phi(q^2)}f_0(0)P(0)\Phi(0),
\label{equation:ff0}
\end{equation}
where the usual kinematic constraint
$f_+(0)=f_0(0)$~\cite{Fajfer:2015ixa} is used in this fit. The
detailed definitions of variables shown in Eqs.~\ref{equation:ffp}
and~\ref{equation:ff0} are given in
Ref.~\cite{supplemental_material}. The Wilson coefficient $c_S^\ell$
is generally allowed to be complex, thus two parameters ${\rm
Re}(c_S^{\mu})$ and ${\rm Im}(c_S^{\mu})$ are introduced for the muon
channels. 
For the positron one, 
the partial decay rates are only
sensitive to the modulus $|c_S^{e}|$ and the forward-backward
asymmetry is not sensitive to the scalar current contribution, thus only
one parameter $|c_S^{e}|$ is considered.

The fit is performed by minimizing the $\chi^2$ function constructed as
\begin{equation}
	\chi^2 = \chi^2_{\Delta \Gamma} + {\chi^2_{A_{{\rm FB}}}}.
\end{equation}
The decay rate part $\chi^2_{\Delta \Gamma}$ is defined as
\begin{equation}
\sum_{i,j}{(\Delta \Gamma_i^{\rm msr}-\Delta \Gamma_i^{\rm fit})(C^{-1}_{\Delta \Gamma})_{ij}(\Delta \Gamma_j^{\rm msr}-\Delta \Gamma_j^{\rm fit})},
\end{equation}
which sums over all four signal channels and takes the correlations between different $q^2$ intervals and signal channels into consideration with the sum of statistical and systematic covariance matrices $C_{ij}=C_{ij}^{\rm stat}+C_{ij}^{\rm syst}$. Here, $\Delta\Gamma^{\rm msr}$ and $\Delta \Gamma^{\rm fit}$ are the measured and fitted decay rates, respectively. The forward-backward asymmetry part $\chi^2_{A_{\rm FB}}$ is defined similarly as
\begin{equation}
\sum_{i,j}{(A_{{\rm FB},i}^{\rm msr}-A_{{\rm FB},i}^{\rm fit})(C^{-1}_{\rm FB})_{ij}(A_{{\rm FB},j}^{\rm msr}-A_{{\rm FB},j}^{\rm fit})},
\end{equation}
where the correlations between different signal channels are ignored, since the only correlated uncertainty caused by the correction factors of $\ell^+$ is limited.

\begin{figure*}[htbp]
	\centering
	\includegraphics[width=18.6cm]{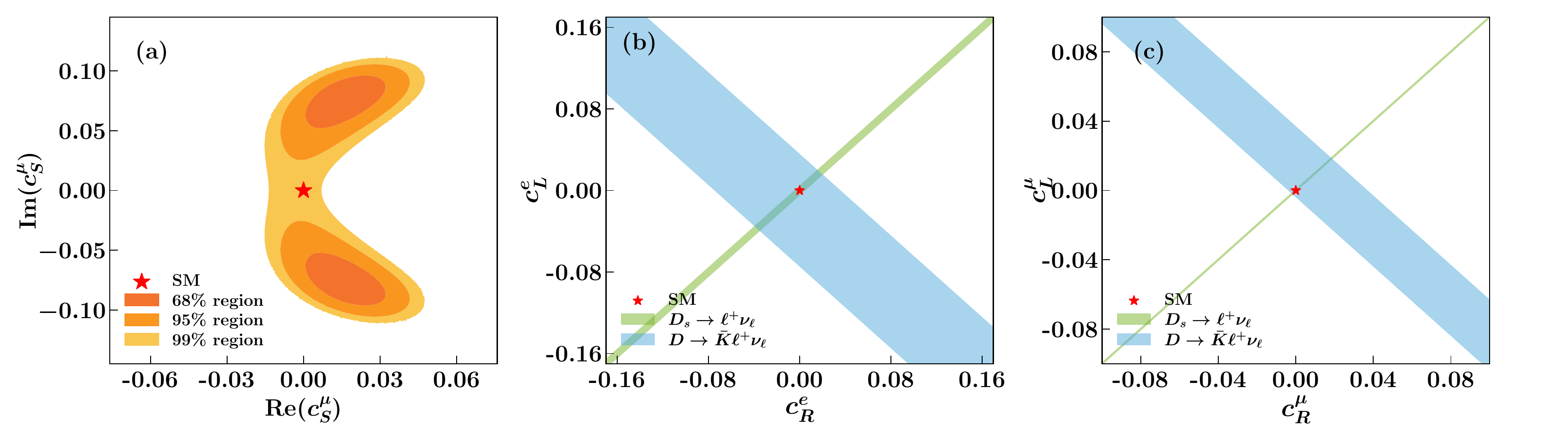}
	\caption{The confidence regions of (a) the scalar combination
	of complex Wilson coefficients $c_S^\mu$ with probabilities of
	68\%, 95\%, and 99\% from $\klnu$, as well as (b)(c) the
	right- and left-handed real Wilson coefficients $c^\ell_{R}$
	and $c^\ell_{L}$ with probabilities of 95\% by combining with the
	$D_s^+\to\ell^+\nu_\ell$ decay rate. The red dot indicates the
	SM value.}
	\label{fig:contour_muon}
\end{figure*}

First, two different fits are performed with and without the
scalar current contribution, with the $c_S^\ell$ shown in
Eq.~\ref{eq:helicity_amp} allowed to float and fixed to zero, respectively. Corresponding fit projections are shown in Fig.~\ref{fig:combined_fit},
and the fitted parameters are summarized in Table~\ref{tab:fited_pars}. No significant deviation from the SM is found
in the positron channels with the fitted $|c^e_S|$ consistent with zero. For the muon channels, $c^\mu_S$ is found to deviate from zero as shown in Fig.~\ref{fig:contour_muon}(a), especially on its imaginary part.

Given the LFU may not hold for the scalar current with possible dependence on lepton mass, its coupling with muon could be significantly larger than that of the positron. Therefore, to estimate the significance of non-zero $c^\mu_S$ alone, the third fit is performed by floating $c^\mu_S$ and fixing $|c^e_S|=0$ as the alternative hypothesis, which yields $\chi^2/{\rm ndf}=135.9/140$. Compared to the null hypothesis without a scalar current with $\chi^2/{\rm ndf}=143.5/142$, the changes $\Delta\chi^2=7.6$ and $\Delta {\rm ndf}=2$ suggest a significance of 2.3$\sigma$ based on the Wilks' theorem.




\begin{table}[htbp]
	\centering
	\setlength\tabcolsep{6pt}
	\caption{Fitted parameters and overall fit qualities of fits
	with and without a scalar current~(SC) contribution, where the
	first uncertainties are statistical and the second are
	systematic.}
	\scalebox{0.92}{
	\begin{footnotesize}
	\begin{tabular}{ccc}
		\hline
		\hline
		Variable & With SC & Without SC \\ \hline
		$f_{+}(0)|V_{cs}|$ & $0.7193\pm0.0008\pm0.0014$ &  $0.7183\pm0.0007\pm0.0014$ \\
		$r_1$ & $-2.23\pm0.04\pm0.02$ &  $-2.28\pm0.02\pm0.02$\\
		$|c_S^{e}|$ & $0.02\pm0.02\pm0.02$ & ---\\
		${\rm Re}(c_S^{\mu})$  & $0.017\pm0.008\pm0.006$ & ---\\
		${\rm Im}(c_S^{\mu})$  & $\pm(0.077\pm0.011\pm0.009)$ & --- \\
		$\chi^2/{\rm ndf}$ & $135.8/139$ & $143.5/142$ \\
		\hline
		\hline
	\end{tabular}
	\end{footnotesize}
	}
	\label{tab:fited_pars}
\end{table}

Further constraints on the right- and left-handed components of the scalar
current are obtained by combining with the decay rates of
$D_s^+\to\ell^+\nu_\ell$, which are sensitive to the pseudoscalar
combination of Wilson coefficients $c^\ell_P=c^{\ell}_R-c^{\ell}_L$
as~\cite{Fajfer:2015ixa}
\begin{equation}
\begin{split}
\mathcal{B}(D_s^+\to\ell^+\nu_\ell)&=\tau_{D_s}\frac{m_{D_s}}{8\pi}f^2_{D_s}\left(1-\frac{m_\ell^2}{m^2_{D_s}}\right)^2 G^2_F \\
&\times|V_{cs}|^2 m^2_{\ell} |1-c^\ell_P\frac{m^2_{D_s}}{(m_c+m_s)m_\ell}|^2.
\end{split}	
\end{equation}
Here, the measured BFs $\mathcal{B}(D_s^+\to \ell^+\nu_\ell)$ are taken
from the Particle Data Group~\cite{ParticleDataGroup:2024cfk}, and
the inputs of $f_{D_s}=249.9\pm0.5$ MeV from a Lattice quantum
chromodynamics~(LQCD) calculation~\cite{Bazavov:2017lyh} and
$|V_{cs}|=0.97349\pm0.00016$ from the global SM
fit~\cite{ParticleDataGroup:2024cfk} are used. Since the
$D_s^+\to\ell^+\nu_\ell$ decay cannot constrain the real and imaginary parts
of $c^\ell_P$ simultaneously, the Wilson coefficient $c^\ell_{R(L)}$
is assumed to be real with T-parity conserved here. As shown in
Figs.~\ref{fig:contour_muon}(b) and~\ref{fig:contour_muon}(c), both
the real parts of the right- and left-handed Wilson coefficients
$c^\ell_{R}$ and $c^\ell_{L}$ are consistent with zero.

Furthermore, the simultaneous fit without a NP contribution gives the
product of the FF $f_+(0)$ and the modulus of the CKM matrix element
$|V_{cs}|$ to be $f_{+}(0)|V_{cs}|=0.7183\pm0.0007_{\rm
	stat}\pm0.0014_{\rm syst}$. Taking $|V_{cs}|=0.97349\pm0.00016$
given by the PDG~\cite{ParticleDataGroup:2024cfk} as input, we
determine $f_+(0)=0.7378\pm0.0007_{\rm stat}\pm0.0014_{\rm syst}$,
which results in a stringent test of various theoretical
calculations. Conversely, taking $f_+(0)=0.7452\pm0.0031$ calculated with LQCD~\cite{FermilabLattice:2022gku}, we obtain
$|V_{cs}|=0.9639\pm0.0009_{\rm stat}\pm0.0018_{\rm
	syst}\pm0.0040_{\rm LQCD}$, which is important for testing the CKM
matrix unitarity.

In summary, by analyzing 20.3 fb$^{-1}$ of $e^+e^-$ collision data
taken at $\sqrt{s}=3.773$ GeV by BESIII, we report the first
measurement of the forward-backward asymmetries as well as the precise
measurements of the partial decay rates of $\klnu$. A simultaneous fit
to these kinematic variables gives the first experimental constraint
on the scalar current contribution in the $\klnu$ transition
($\ell=e,\mu$). The complex Wilson coefficient is determined to be ${\rm
Re}(c_S^{\mu})=0.017\pm0.008_{\rm stat}\pm0.006_{\rm syst}$ and
${\rm Im}(c_S^{\mu})=\pm(0.077\pm0.011_{\rm stat}\pm0.009_{\rm
syst})$, which deviates from the SM by 2.3$\sigma$. This result
provides an important constraint on the scalar current contribution in the
SL charmed meson decays, further provides new insights into the
anomalies observed in the beauty
sector~\cite{BaBar:2012obs,Belle:2015qfa,LHCb:2023zxo,LHCb:2023uiv,LHCb:2024jll,LHCb:2013ghj,Belle:2016fev,CMS:2017rzx,ATLAS:2018gqc,LHCb:2024onj},
and facilitates the search for new scalar bosons beyond the Standard
Model under various new-physics scenarios. Besides, the improved measurements of the BFs and tests of LFU in $\klnu$ decays, the hadronic transition FFs $f_{+}(0)$ as well as the CKM matrix element $|V_{cs}|$, which are in the highest precision to date, are also presented.

The BESIII Collaboration thanks the staff of BEPCII (https://cstr.cn/31109.02.BEPC) and the IHEP computing center for their strong support. This work is supported in part by National Key R\&D Program of China under Contracts Nos. 2023YFA1606000, 2023YFA1606704; National Natural Science Foundation of China (NSFC) under Contracts Nos. 11635010, 11935015, 11935016, 11935018, 12025502, 12035009, 12035013, 12061131003, 12192260, 12192261, 12192262, 12192263, 12192264, 12192265, 12221005, 12225509, 12235017, 12342502, 12361141819; the Chinese Academy of Sciences (CAS) Large-Scale Scientific Facility Program; the Strategic Priority Research Program of Chinese Academy of Sciences under Contract No. XDA0480600; CAS under Contract No. YSBR-101; 100 Talents Program of CAS; The Institute of Nuclear and Particle Physics (INPAC) and Shanghai Key Laboratory for Particle Physics and Cosmology; ERC under Contract No. 758462; German Research Foundation DFG under Contract No. FOR5327; Istituto Nazionale di Fisica Nucleare, Italy; Knut and Alice Wallenberg Foundation under Contracts Nos. 2021.0174, 2021.0299, 2023.0315; Ministry of Development of Turkey under Contract No. DPT2006K-120470; National Research Foundation of Korea under Contract No. NRF-2022R1A2C1092335; National Science and Technology fund of Mongolia; Polish National Science Centre under Contract No. 2024/53/B/ST2/00975; STFC (United Kingdom); Swedish Research Council under Contract No. 2019.04595; U. S. Department of Energy under Contract No. DE-FG02-05ER41374.

\bibliography{reference.bib}

\end{document}